# Effect of Ge substitution for Si on the anomalous magnetocaloric and magnetoresistance properties of $GdMn_2Si_2$ compounds

Pramod Kumar, Niraj K. Singh and K. G. Suresh[a]

*Department of Physics, I. I. T. Bombay, Mumbai 400076, India*

A. K. Nigam and S. K. Malik

*Tata Institute of Fundamental Research, Homi Bhabha Road, Mumbai 400005, India*

The effect of Ge substitution on the magnetization, heat capacity, magnetocaloric effect and magnetoresistance of $GdMn_2Si_{2-x}Ge_x$ (x=0, 1, and 2) compounds has been studied. The magnetic transition associated with the Gd ordering is found to change from second order to first order on Ge substitution. Magnetic contributions to the total heat capacity and the entropy have been estimated. Magnetocaloric effect has been calculated in terms of adiabatic temperature change ($\Delta T_{ad}$) as well as isothermal magnetic entropy change ($\Delta S_M$) using the heat capacity data. The temperature dependence of the magnetocaloric effect in all the three compounds have shown broad peaks. The maximum values of $\Delta S_M$ and $\Delta T_{ad}$ for $GdMn_2Ge_2$ are found to be 5.9 J/kgK and 1.2 K, respectively. The magnetoresistance is found to be very large and positive with a maximum value of about 22% in the case of $GdMn_2Ge_2$. In the other two compounds also, the magnetoresistance is predominantly positive, except in the vicinity of the Gd ordering temperature. The anomalous nature of the magnetocaloric effect and the magnetoresistance has been attributed to the canted magnetic structure of these compounds.

--------------------------------------------------------------------------------

[a] Corresponding author (Electronic mail: suresh@phy.iitb.ac.in)

## I. INTRODUCTION

Magnetocaloric effect (MCE), which manifests as the heating or cooling of magnetic materials due to a varying magnetic field, has attracted considerable attention recently [1-4]. The origin of MCE in many materials has been explained and its practical use to achieve low temperatures is being anticipated. Giant magnetocaloric effect exhibited by many rare earth (R)–transition metal (TM) intermetallic compounds render them as potential refrigerants for magnetic refrigerators. Large value of MCE spread over a wide temperature range is considered as one of the most important requirements of a practical magnetic refrigerant system. The compounds such as $Gd_5Si_2Ge_2$ which show field-induced magnetic transitions and/or structural transitions have been found to exhibit giant magnetocaloric effect[5,6] A few transition metal based compounds also show giant magnetocaloric effect, the representative examples being MnAs and $LaFe_{13}$ based systems, which undergo a first-order ferromagnetic to paramagnetic transition at their ordering temperature[7-10].

Among the $R$-TM intermetallics, $RMn_2X_2$ (X=Si, Ge) have attracted a lot of attention due to their interesting magnetic properties [11-14]. These compounds, in general, crystallize in the tetragonal $ThCr_2Si_2$-type structure. Compounds of this series exist in layered structure with the sequence Mn-X-R-X-Mn, with the layers perpendicular to the c-axis. It is found that both R and Mn ions possess magnetic moments in these compounds. The inter layer and the intra layer Mn-Mn exchange interactions are very sensitive to the intra layer Mn-Mn distance ($d_{Mn-Mn}^a$). It has been reported that if $d_{Mn-Mn}^a$ >2.87 Å, the intra layer Mn-Mn coupling is antiferromagnetic and the inter layer Mn-Mn coupling is ferromagnetic. When 2.84 Å< $d_{Mn-Mn}^a$ <2.87 Å, both the intra and the inter layer couplings are antiferromagnetic. For $d_{Mn-Mn}^a$ <2.84, there is effectively no intra-layer spin component and the inter layer coupling remains antiferromagnetic. The antiferromagnetic ordering of the inter-layer Mn sublattice, in general, exists above the room temperature, while R sublattice remains disordered. For example, the Neel temperatures of the Mn sublattice in GdMn$_2$Si$_2$ and GdMn$_2$Ge$_2$ are 465 K and 365 K, respectively[11,12]. As the temperature is reduced, the R moments also order magnetically at T$_C^R$, below which Mn and R sublattices couple ferromagnetically or antiferromagnetically depending on whether the rare earth is light or heavy. Another interesting behavior of this series is that, while the $R$Mn$_2$Si$_2$ compounds show second order magnetic transition (SOT) associated with the rare earth ordering, the iso-structural $R$Mn$_2$Ge$_2$ compounds show first order magnetic transition (FOT)[15]. The occurrence of FOT is expected to result in considerable MCE and magnetoresistance (MR).

Another advantage with this series is the possibility of tuning the magnetic ordering temperature with the help of substitutions at the rare earth and Si/Ge sites. In the search of 'table-like' MCE, it is of interest to study the variation of MCE in these substituted compounds. As part of our investigations on the RMn$_2$X$_2$ compounds, we have recently reported the magnetic and magnetocaloric studies of Gd$_{1-x}$Sm$_x$Mn$_2$Si$_2$ compounds[16]. In this paper, we report the effect of Ge substitution for Si on the magnetic, magnetocaloric and magnetoresistance properties of GdMn$_2$Si$_{2-x}$Ge$_x$ with x=0, 1 and 2. The main aim of the work is to study the variation of MCE and MR as the nature of the magnetic transition at T$_C^R$ (=T$_C^{Gd}$) changes from second order (in x=0) to first order (in x=2).

## II. EXPERIMENTAL DETIALS

Polycrystalline samples of GdMn$_2$Si$_{2-x}$Ge$_x$ with x=0, 1 and 2 were synthesized by arc melting the constituent elements in stochiometric proportions in a water-cooled copper hearth under purified argon atmosphere. The purity of the starting elements was 99.9% for Gd and 99.99 % for Mn, Si and Ge. The ingots were melted several times to ensure homogeneity. The as-cast samples were characterized by room temperature powder x-ray diffractograms (XRD), collected using Cu-K$_\alpha$ radiation. The magnetization (M) measurements, both under 'zero-field-cooled' (ZFC) and 'field-cooled' (FC) conditions, in the temperature (T) range of 5-150 K and upto a maximum field (H) of 80 kOe were performed using a vibrating sample magnetometer (VSM, OXFORD instruments). In the ZFC mode, the samples were cooled in the absence of a field and the magnetization was measured during warming, by applying a nominal field of 200 Oe. For the FC measurement, the samples were cooled in presence of a field and the magnetization data was collected during warming, under the same field (200 Oe). The heat capacity (C) and electrical resistivity (ρ) were measured in the temperature range of 2–280 K and in fields up to 50 kOe, using a physical property measurement system (PPMS, Quantum Design). The heat capacity was measured using the relaxation method and the electrical resistivity was measured by employing the linear four-probe technique. The magnetocaloric effect has been calculated both in terms of isothermal magnetic entropy change and adiabatic temperature change, using the heat capacity data in zero field and at 50 kOe.

**III. RESULTS AND DISCUSSION**

The Rietveld refined XRD patterns of GdMn$_2$Si$_{2-x}$Ge$_x$ compounds at room temperature are shown in Fig. 1. The difference between the experimental and the calculated intensities is given at the bottom of each plot. The refinement of these patterns confirms that all the compounds are single phase with the ThCr$_2$Si$_2$ structure (space group=I4/mmm). The lattice parameters and the Mn-Mn bond lengths along the *a*-axis and the *c*-axis ($d^a_{Mn-Mn}$ and $d^c_{Mn-Mn}$), as obtained from the refinement, are given in Table I. It may be noticed from the table that $d^a_{Mn-Mn}$ and $d^c_{Mn-Mn}$ increase with increase in Ge content. This variation is attributed to the larger ionic radius of Ge compared to that of Si. Furthermore, it may be noticed from the table that $d^a_{Mn-Mn}$ for the compounds with x=0 and 1 is less than the critical value (2.84 Å) needed for the existence of intra-layer spin component of the Mn sublattice, whereas it is just above the critical value in the case of x=2.

Fig. 2 shows the temperature dependence of the magnetization of GdMn$_2$Si$_{2-x}$Ge$_x$ compounds collected in an applied field of 200 Oe, under ZFC and FC conditions. The sharp decrease in the magnetization seen in these compounds is attributed to the magnetic order-disorder transition occurring in the Gd sublattice. The corresponding transition temperature is termed as $T_C^{Gd}$ and is calculated from the d$M$/d$T$ vs. T plot. The variation of $T_C^{Gd}$ with $x$ is also shown in Table I. It is noticed that the sharpness of the transition increases with Ge concentration. The width of the transition has been calculated from the full width at half maxima of the d$M$/d$T$ vs. T plots and found that the values are 2.2 K, 0.9 K and 0.2 K for x=0, 1 and 2, respectively. It can also be seen from Fig. 2 that the magnetization in these compounds is almost zero above 100 K. The $T_C^{Gd}$ for the compounds with x= 0, 1 and 2 are found to be 63 K, 57 K and 95 K, respectively. It is to be noted that the $T_C^{Gd}$ values in the compounds with x=0 and x=2 are in good agreement with the reported values[17]. It is interesting to note that $T_C^{Gd}$ values in this series is almost constant for x= 0 and 1, whereas it shows considerable increase in x=2. It may be recalled here that $d_{Mn-Mn}^a$ for the compounds with x= 0 and 1 is less than the critical value required for the occurrence of intra-layer Mn spin component whereas it is larger than the critical value in the compound with x=2. Therefore, according to the spacing conditions mentioned earlier, the compound with x=2 would have a nonzero intra-layer Mn spin component. This may cause an overall increase in the R-TM exchange strength and this may be the reason for the higher $T_C^{Gd}$ in this compound. It is also clear from Fig. 2 that the temperature variation of ZFC and FC magnetization data of the compounds with x= 0 and 1 show negligibly small thermomagnetic irreversibility while the irreversibility is considerable in x=2.

Fig. 3 shows the magnetization isotherms of GdMn$_2$Si$_{2-x}$Ge$_x$ compounds at 5 K. It can be seen from the figure that the magnetic moment, for an applied field of 80 kOe, of the compounds with x= 0 and 1 is about 4.5 $\mu_B$/f.u. and 4.9 $\mu_B$/f.u., respectively, whereas it is 3.1 $\mu_B$/f.u for the compound with x= 2. It may be mentioned here that in the RMn$_2$X$_2$ compounds, below the rare earth ordering temperature, the Mn moments order ferrimagnetically (ferromagnetically) for heavy (light) rare earths. Therefore, the reduction in the magnetization for the compound with x=2 suggests an increase in the Mn moment as x is increased from 0 to 2. This observation is consistent with the increase in the $T_C^{Gd}$ of the compounds with x=2, as mentioned earlier.

Fig. 4 shows the Arrott plots of GdMn$_2$SiGe and the inset shows the temperature variation of the spontaneous magnetization calculated from the Arrott plots (solid circles). The temperature variation of the spontaneous magnetization has been fitted to the Stoner model (dotted line) and the spin fluctuation model (solid line), i.e. to the relations $M_S(T) = M_s(0)\left(1-\left(\frac{T}{T_C^{Gd}}\right)^2\right)^{\frac{1}{2}}$ and $M_S(T) = M_s(0)\left(1-\frac{T}{T_C^{Gd}}\right)^{\frac{1}{2}}$ respectively[18]. Here M$_s$ refers to the saturation magnetization. As can be seen, the spin fluctuation model gives a better fit than the Stoner model. The spin fluctuation model fit yielded the value of $T_C^{Gd}$ to be 50.8 K and M$_s$(0) to be 67.6 emu/g. The value of $T_C^{Gd}$ thus obtained is in agreement with that determined from the low field M-T plot (Fig. 2).

Fig.5 shows the temperature variation of heat capacity of GdMn$_2$Si$_2$ under zero field as well as under 10 and 50 kOe. The λ-like peak observed in the zero field heat capacity can be clearly seen from the expanded plot shown in the inset of this figure. This type of behavior confirms that this compound undergoes a second order phase transition at $T_C^{Gd}$=63 K. With increasing magnetic field, the peak becomes broader and shifts to higher temperatures. Finally, the peak disappears when the field is increased to 50 kOe. A similar observation has been made in the compound with x=1 as well. However, the C-T plot of GdMn$_2$Ge$_2$ (Fig. 6c) shows that the peak at $T_C^{Gd}$=95 K is very sharp in zero field. In all the compounds, the C-T plots have steep slopes at around room temperature, due to the magnetic ordering of the Mn sublattice.

In general, the heat capacity of metallic magnetic systems can be written as
$$C_{tot}=C_{ph}+C_{el}+C_M \text{----------(1)}$$
where C$_{el}$, C$_{ph}$ and C$_M$ are the electronic, lattice and magnetic contributions respectively. In order to separate C$_M$ from C$_{tot}$, the first two terms of eqn. (1) have to be evaluated. This has been done, by taking the iso-structural nonmagnetic counterparts namely LaFe$_2$Ge$_2$ and LaFe$_2$Si$_2$, whose heat capacity can be written as

$$C_{tot} = \gamma T + 9NR(T/\theta_D)^3 \int_0^{\theta_D/T} \frac{x^4 e^x}{(e^x-1)^2} dx \quad \text{------(2)}$$

where the first term represents the electronic contribution and the second term corresponds to the phonon contribution. Here N is the number of atoms per formula unit (N=5 in this case), R is the molar gas constant, $\gamma$ is the electronic coefficient and $\theta_D$ the Debye temperature. It has been reported[19,20] that the heat capacity of $LaFe_2Ge_2$ and $LaFe_2Si_2$ follows the above expression satisfactorily with an effective $\theta_D$= 180K and 280K and $\gamma$=36 mJ/mol K and 22.7mJ/mol K, respectively. By using the relation[19]

$$M_{GdMn_2(Si,Ge)_2} \left(\theta_{GdMn_2(Si,Ge)_2}\right)^2 = M_{LaFe_2(Si,Ge)_2} \left(\theta_{LaFe_2(Si,Ge)_2}\right)^2 \quad \text{-----(3)}$$

we have calculated the $\theta_D$ values to be 263 K, 232 K and 171 K for $GdMn_2Si_{2-x}Ge_x$ compounds with x=0, 1 and 2 respectively. Using these values, the nonmagnetic contribution to the heat capacity has been calculated and are shown in Fig. 6a-c (solid lines). The magnetic specific heat $C_M$, is obtained by subtracting the nonmagnetic contribution from the total heat capacity. The temperature dependence of $C_M$ is also shown in Fig. 6a-c (open circles). The value of $C_M$ at $T_C^{Gd}$ is about 12 J/molK in $GdMn_2Si_2$ and $GdMn_2SiGe$ compounds and 180J/molK in $GdMn_2Ge_2$. The large value of $C_M$ in $GdMn_2Ge_2$ is due to the first order magnetic transition occurring at $T_C^{Gd}$. In all the compounds, $C_M$ increases with increasing temperature. The insets of the Fig. 6a-c show the variation of the magnetic entropy ($S_M$) with temperature. At $T_C^{Gd}$, the values of magnetic entropy are 13.1 J/mol K for x=0 &1 and 15.1 J/mol K for the compound with x= 2. The entropy jump of 3.3 J/molK at $T_C^{Gd}$ in $GdMn_2Ge_2$ matches well with the theoretical value of 3.7J/molK, obtained by Wada et al.[21] It is of interest to note that the Rln(2J+1) value for $Gd^{3+}$ (J=7/2) ion is 17.2J/mol K. In all the compounds, at high temperatures, the $S_M$ versus T plots shows a convex type curvature.

The magnetic entropy below $T_C^{Gd}$ mainly originates from the Gd-sublattice. The full entropy of the Gd moment is released at $T_C^{Gd}$, above which the entropy of the Gd sublattice is almost temperature independent. On the other hand, the Mn sublattice contributes to $S_M$ mainly at temperatures above $T_C^{Gd}$. Therefore, the curvature seen in the $S_M$ vs. T curves is due to the Mn sublattice, whose antiferromagnetic ordering decreases with temperature.

Fig.7 shows the heat capacity below the $T_C^{Gd}$ of all the three compounds, fitted to the relation $C = aT^{-\frac{1}{2}} \exp(-\frac{\Delta}{T})$, where $\Delta$ is the minimum energy required to excite a spin wave in the anisotropy field[22]. As shown in the figure, the value of $\Delta$ is the same for x=0 and 1, whereas it increases considerably in x=2. As mentioned earlier, due to the increase in the lattice parameter, the Gd-Mn exchange interaction increases with Ge concentration. The increase in the Gd-Mn exchange interaction may be the reason for the increase in $\Delta$ with Ge concentration.

The magnetocaloric effect of $GdMn_2Si_{2-x}Ge_x$ compounds has been calculated both in terms of isothermal entropy change ($\Delta S_M$) as well as adiabatic temperature change ($\Delta T_{ad}$). This has been done using the methods reported earlier[16,23,24]. Fig. 8a-c show the temperature variation of $\Delta S_M$ and $\Delta T_{ad}$ for a field change ($\Delta H$) of 50kOe, for all compounds. $\Delta S_M$ values calculated from the C-H-T data were found to be in close agreement with those obtained from the M-H-T data. It can be seen from Fig. 8a-b that $\Delta S_M$ vs. T plots of $GdMn_2Si_2$ and $GdMn_2SiGe$ show a broad maximum, unlike the general observation of a sharp peak in many materials. The magnetic entropy change seems to be quite considerable even at temperatures well below $T_C^{Gd}$. In the case of $GdMn_2Ge_2$, the low temperature value of the entropy change has increased to such an extent that it has resulted in a 'table-like' behavior with a very sharp peak near $T_C^{Gd}$, as shown in Fig. 8c. The MCE behavior seen in this compound is similar to that of Ho(Ni,Fe)$_2$ compounds[24]. The maximum values of $\Delta S_M$ ($\Delta S_M^{max}$), for $GdMn_2SiGe$ and $GdMn_2Ge_2$ are found to be 6.3 J/kg K and 5.9 J/kg K at about 57 and 96K, respectively. It is also evident from Figs. 8a-c that the temperature dependence of $\Delta T_{ad}$ is identical to that of $\Delta S_M$.

In this context, it is of interest to compare the MCE variation in the iso-structural $GdRu_2Ge_2$ reported by Tegus et al.[25] MCE of this compound calculated using the magnetization and heat capacity data shows a sharp peak near the magnetic ordering temperature of the Gd sublattice. Since Ru is nonmagnetic in this compound, it is reasonable to attribute the differences in the MCE behavior in GdMn$_2$Si$_{2-x}$Ge$_x$ series to the magnetic ordering associated with the Mn sublattice. The anomalous shape of the MCE plots in GdMn$_2$Si$_{2-x}$Ge$_x$ compounds may be explained by taking into account the magnetic interactions of Gd and Mn sublattices. Based on the magnetoresistance data, Wada et al.[26] have recently suggested a new type of magnetic state for GdMn$_2$Ge$_2$ compound at temperatures below $T_C^{Gd}$. Since Gd-Mn coupling is ferrimagnetic, with increase in applied field, the net field acting on the Mn moments decreases and as a result, the Mn sublattice tends to become antiferromagnetically coupled. Consequently, a canted magnetic structure develops at temperatures below $T_C^{Gd}$ in presence of an applied field. Fujiwara et al.[13] have also reported a canted magnetic structure for RMn$_2$X$_2$ compounds. We feel that this canted structure may be responsible for the unusually large magnetic entropy change at temperatures well below $T_C^{Gd}$, resulting in broad MCE peaks in the compounds with x=0 and 1. Since the Mn sublattice moment is more in the compound with x=2, as compared to the case of x=0 and 1, the effects of canting will be more in the former case. This may be the reason for the 'table-like' MCE. The sharp peak near $T_C^{Gd}$ is due to the first order nature of the magnetic transition. It should also be noted that in all the three compounds, the absolute values of $\Delta S_M(T_C^{Gd})$ are much smaller than the theoretical magnetic entropy of Gd given by Rln(2J+1)=17 J/kg K (with J=7/2).

Another feature that can be seen from Fig. 8c is the negative MCE at temperatures above $T_C^{Gd}$=95 K in the case of GdMn$_2$Ge$_2$, indicating the predominantly antiferromagnetic nature of the Mn sublattice in that rage of temperature. A similar observation has been made in GdMn$_2$Si$_2$ at temperatures well above $T_C^{Gd}$ (Fig. 8a). It should also be noted that the Neel temperatures of the Mn sublattice in GdMn$_2$Ge$_2$ and GdMn$_2$Si$_2$ are 365 K and 465 K.

In order to find out the dependence of MCE on the applied field, we have plotted the $\Delta S_M^{max}$ values as a function of $H^2$ at temperatures above $T_C^{Gd}$ and the representative plots for GdMn$_2$SiGe and GdMn$_2$Ge$_2$ are shown in Fig. 9a-b. It can be seen that at temperatures just above $T_C^{Gd}$, the quadratic dependence is not very good, but the fit improves considerably at higher temperatures. A quadratic dependence of MCE on the field is indicative of the presence of spin fluctuations, as reported by Das and Rawat.[27] Very recently Kumar et al[16] and Tripathy et al.[28] have also observed the same dependence of MCE in $Gd_{1-x}Sm_xMn_2Si_2$ and $Gd_3(Co,Ni)$ compounds. Therefore, the quadratic dependence seen in the present series of compounds may also be indicative of the role of spin fluctuations on the magnetic entropy change at temperatures above $T_C^{Gd}$. The spin fluctuations may arise from the rare earth moments and also from the Mn sublattice.

Figs. 10a-c show the temperature variation of electrical resistivity of $GdMn_2Si_{2-x}Ge_x$ compounds in zero field and 50 kOe. It may be noted that all the compounds show metallic behavior. The residual resistivity is high, probably due to the large number of atoms in the unit cell as well as the presence of micro-cracks in the samples. Though no anomalies could be clearly observed in the vicinity of $T_C^{Gd}$, they are prominently seen in the $d\rho/dT$ plots shown in Fig. 11. As can be seen, Gd ordering is marked by a broad maximum in $GdMn_2Si_2$, while it is marked by a sharp peak in $GdMn_2Ge_2$. The order of magnetic transition may be responsible for this difference. The broad maximum seen at temperatures above $T_C^{Gd}$, especially in $GdMn_2Si_2$ can be explained on the basis of the short range fluctuations of the s-f exchange parameter[29]. A similar behavior has been reported in many materials such as $GdNi_2$, $GdRh_2$ and $GdPt_2$[30].

According to Matheissen's rule, the resistivity of a ferromagnetic metal could be written as $\rho(T)=\rho_0+\rho_{ph}(T)+\rho_{mag}(T)$, where $\rho_0$ is the residual resistivity which is temperature independent. $\rho_{ph}$ and $\rho_{mag}$ are the contributions from the electron-phonon and electron-spin wave scatterings, respectively. Depending on the temperature, $\rho_{ph}$ or $\rho_{mag}$ dominates the total resistivity. In order to find out the dominant contribution, the resistivity data has been analyzed separately, in low and high temperature regions. In magnetically ordered materials, the temperature dependent part of the electrical resistivity arises due to the electron-phonon scattering, the electron-spin wave scattering and electron-electron scattering. In general, at low temperature, the contribution from the electron-phonon scattering is found to be small as compared to the other two contributions[31]. The electrical resistivity due to the electron-electron and electron spin wave scatterings follow a quadratic behavior. As can be seen from the upper inset in Figs.10a-c, the electrical resistivity at low temperature shows a $T^2$ dependence. The relation used for the least square fit can expressed as $\rho(T)=\rho_0+AT^2$. The value of 'A' suggests whether the electron-electron scattering or the electron-spin wave scattering is dominant. In the former case, the value of 'A' would be small (of the order of $10^{-2}\,n\Omega cm/K^2$), whereas 'A' will be quite high (of the order of a few $n\Omega cm/K^2$) in the latter case. In the case of GdMn$_2$Si$_{2-x}$Ge$_x$ compounds the typical values of 'A' is found to be about $2000\,n\Omega cm/K^2$, suggesting that the electron spin-wave scattering is the dominant factor in determining the electrical resistivity in the temperature range of 5-15K. This implies that for these compounds, the spin wave dispersion relation must be quadratic, i.e. $\varepsilon^2(k)=\Delta^2+Dk^2$ where $k$ is spin wave vector and $\Delta$ represents the minimum energy required to excite a spin wave in the anisotropy field, as mentioned earlier.

On the other hand, at high temperatures, the electron-phonon scattering is found to be dominant and this is seen as a linear increases in resistivity with temperature. However deviations from this behavior are possible and these have been attributed to the s-d scattering[31]. In most of the materials, the s-d scattering contribution gives rise to a dependence of the form $\rho(T)=\rho_0+BT-CT^3$. The lower insets of Figs. 10 a-c show the resistivity in the temperature range of 100-300K, fitted to this expression. A good fit is obtained as shown by the solid lines in these figures. The fact that 'C' is positive implies a negative curvature of the resistivity curves in the high temperature range. Such a behavior is generally known to arise from the s-d scattering.

The magnetoresistance (MR), defined as $\frac{\Delta\rho}{\rho}(\%)=\frac{[\rho(H)-\rho(0)]}{\rho(0)}\times 100$, for all the compounds has been studied in various temperature regions, as shown in Fig 12. It can be seen from the figure that the MR for the compounds with x=0 and x=1 is positive in almost the entire temperature range, except in the vicinity of $T_C^{Gd}$. At the lowest temperature, the MR is about 4 % in both these compounds. In contrast, the MR in the compound with x=2 is positive in the entire temperature range with a maximum value of ~22% at 5 K. The magnitude of MR decreases with temperature and at $T_C^{Gd}$ it shows a discontinuous reduction to about 10%, which is almost retained upto to room temperature. This observation seen in GdMn$_2$Ge$_2$ is similar to the one reported by Wada et al.[26] In SmMn$_2$Ge$_2$ also, positive MR has been observed, which is attributed to the Lorentz force term and the spin dependent scattering[32]. Since the Lorentz force term cannot be very much different in the three compounds studied in the present case, we feel that the canted magnetic structure, mentioned earlier, may be responsible for the large positive MR. As in the case of MCE, larger effect of canting by virtue of additional moments on Mn sublattice may be responsible for the large positive MR in GdMn$_2$Ge$_2$. The discontinuous reduction of MR at $T_C^{Gd}$ in this compound must be due to the occurrence of FOT.

## IV. CONCLUSIONS

We have studied the effect of Ge substitution on the structural, magnetic, magneto-thermal and magneto-transport behavior of GdMn$_2$Si$_{2-x}$Ge$_x$ compounds. Powder x-ray diffraction study has shown that the crystal structure remains unchanged upon Ge substitution, but the lattice parameters increase with Ge concentration. The change in the lattice parameters is found to be correlated with the magnetic properties. Magnetization measurements reveal that the magnetic transition associated with the Gd ordering changes from second order to first order with Ge substitution. From the heat capacity study, the magnetic contributions to the total heat capacity and the magnetic entropy have been estimated. Magnetocaloric effect has been calculated in terms of adiabatic temperature change as well as isothermal magnetic entropy change. Contrary to our expectation, a large MCE could not be achieved in GdMn$_2$Ge$_2$, in spite of having FOT. Furthermore, the MCE peaks in all the three compounds are quite broad, indicating significant contributions to the magnetic entropy change at temperatures well below the Gd ordering temperature. The dominant contributions to the electrical resistivity have been identified at different temperature ranges, by fitting the temperature dependence of the resistivity data. The magnetoresistance is found to be very large and positive with a maximum value of about 22% in the case of GdMn$_2$Ge$_2$. In the other two compounds also, the magnetoresistance is predominantly positive, except in the vicinity of the Gd ordering temperature. The anomalous MCE and MR behavior is attributed to the canted magnetic structure.

**Acknowledgement**

One of the authors (KGS) thanks ISRO, Govt. of India for supporting this work through a sponsored research grant.

Table I Lattice parameters (*a* & *c*), unit cell volume (*V*), Mn-Mn bond lengths ($d_{Mn\text{-}Mn}$) and the Gd ordering temperature ($T_C^{Gd}$) in GdMn$_2$Si$_{2-x}$Ge$_x$ compounds

| x | a (Å) | c (Å) | V (Å$^3$) | $d_{Mn-Mn}^c$ (Å) | $d_{Mn-Mn}^a$ (Å) | $T_C^{Gd}$ (K) |
|---|---|---|---|---|---|---|
| 0 | 3.948 | 10.475 | 163.320 | 3.294 | 2.791 | 63 |
| 1 | 3.975 | 10.622 | 167.839 | 3.322 | 2.811 | 57 |
| 2 | 4.026 | 10.881 | 176.361 | 3.389 | 2.847 | 95 |

LIST OF FIGURES

Fig.1 Rietveld refined powder x-ray diffractograms of GdMn$_2$Si$_{2-x}$Ge$_x$ compounds. The plot at the bottom (Blue solid line) shows the difference between the calculated(Red solid line) and experimental(Open circules) patterns in each case.

Fig.2 Temperature variation of magnetization of GdMn$_2$Si$_{2-x}$Ge$_x$ compounds in a field of 200 Oe under ZFC and FC conditions.

Fig.3. M-H isotherms of GdMn$_2$Si$_{2-x}$Ge$_x$ compounds at 5 K.

Fig.4 Arrott plots at different temperatures for GdMn$_2$SiGe compound. The inset shows the variation of spontaneous magnetization obtained from the Arrott plots fitted to the Stoner model (dotted line) and the spin fluctuation model (solid line).

Fig.5 Temperature variation of heat capacity of GdMn$_2$Si$_2$ in 0, 20 and 50 kOe. The inset shows the variation near the rare earth ordering temperature.

Fig.6a-c Temperature variation of heat capacity (C$_{exp}$) of GdMn$_2$Si$_{2-x}$Ge$_x$ in zero applied field (filled squares). The solid line represents the theoretically calculated electronic and phonon contributions (C$_{th}$). The open circles represent the magnetic contribution (C$_M$). The inset shows the magnetic entropy (S$_M$) variation.

Fig. 7 Variation of heat capacity in the low temperature range.

Fig.8a-c Temperature variation of isothermal magnetic entropy change and adiabatic temperature change in GdMn$_2$Si$_{2-x}$Ge$_x$ compounds for a field change of 50kOe.

Fig.9 Variation of isothermal magnetic entropy change as a function of H$^2$ in (a) GdMn$_2$SiGe and (b) GdMn$_2$Ge$_2$.

Fig.10a-c Temperature variation of electrical resistivity of $GdMn_2Si_{2-x}Ge_x$ compounds. The upper inset shows the low temperature data fitted to the relation $\rho(T)=\rho_0+AT^2$. The lower inset shows the data in the range 100-300 K fitted to the relation $\rho(T)=\rho_0+BT-CT^3$.

Fig.11 $d\rho/dT$ vs. T plots of $GdMn_2Si_{2-x}Ge_x$ compounds.

Fig.12 Temperature variation of magnetoresistance of $GdMn_2Si_{2-x}Ge_x$ compounds.

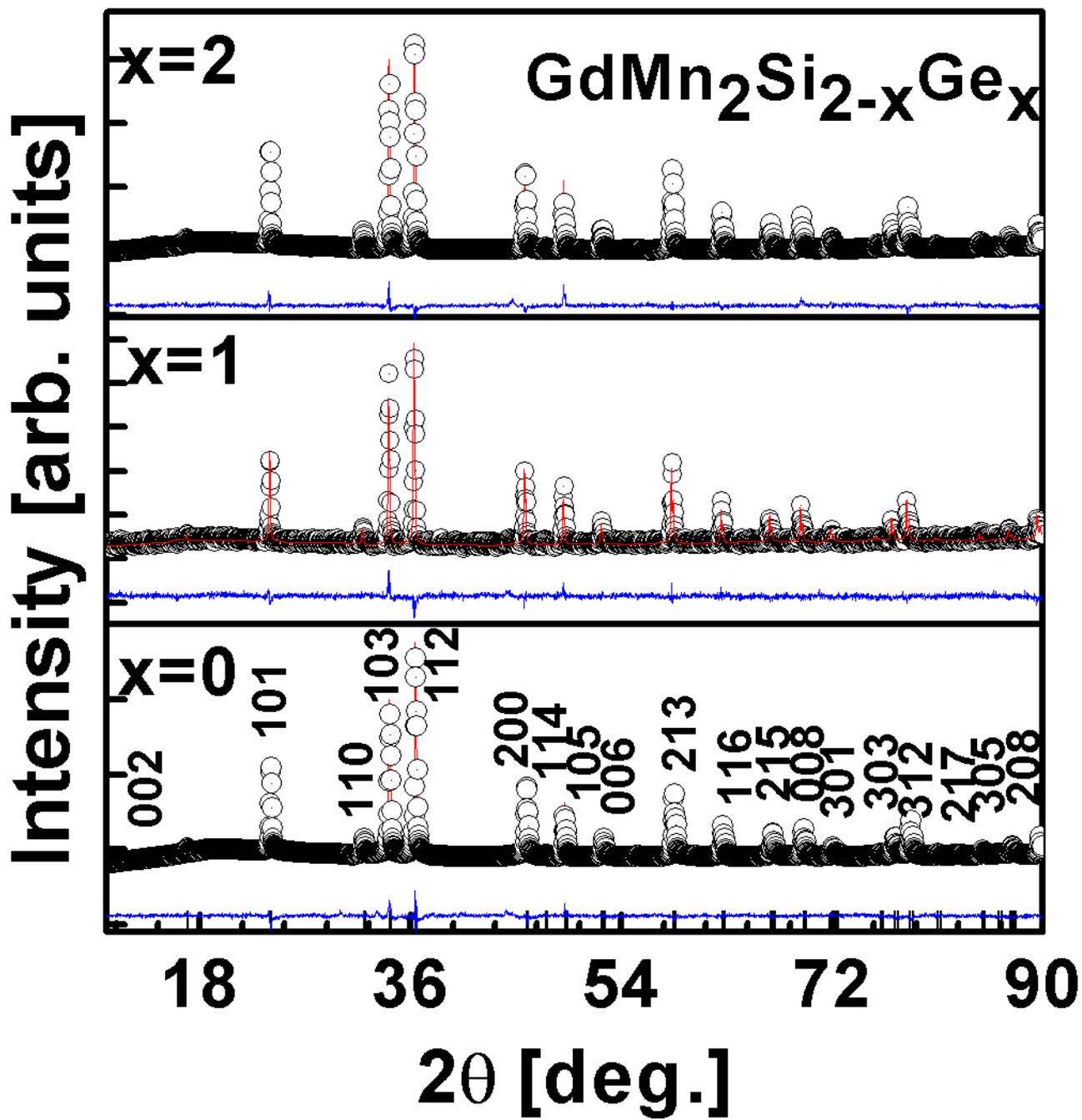

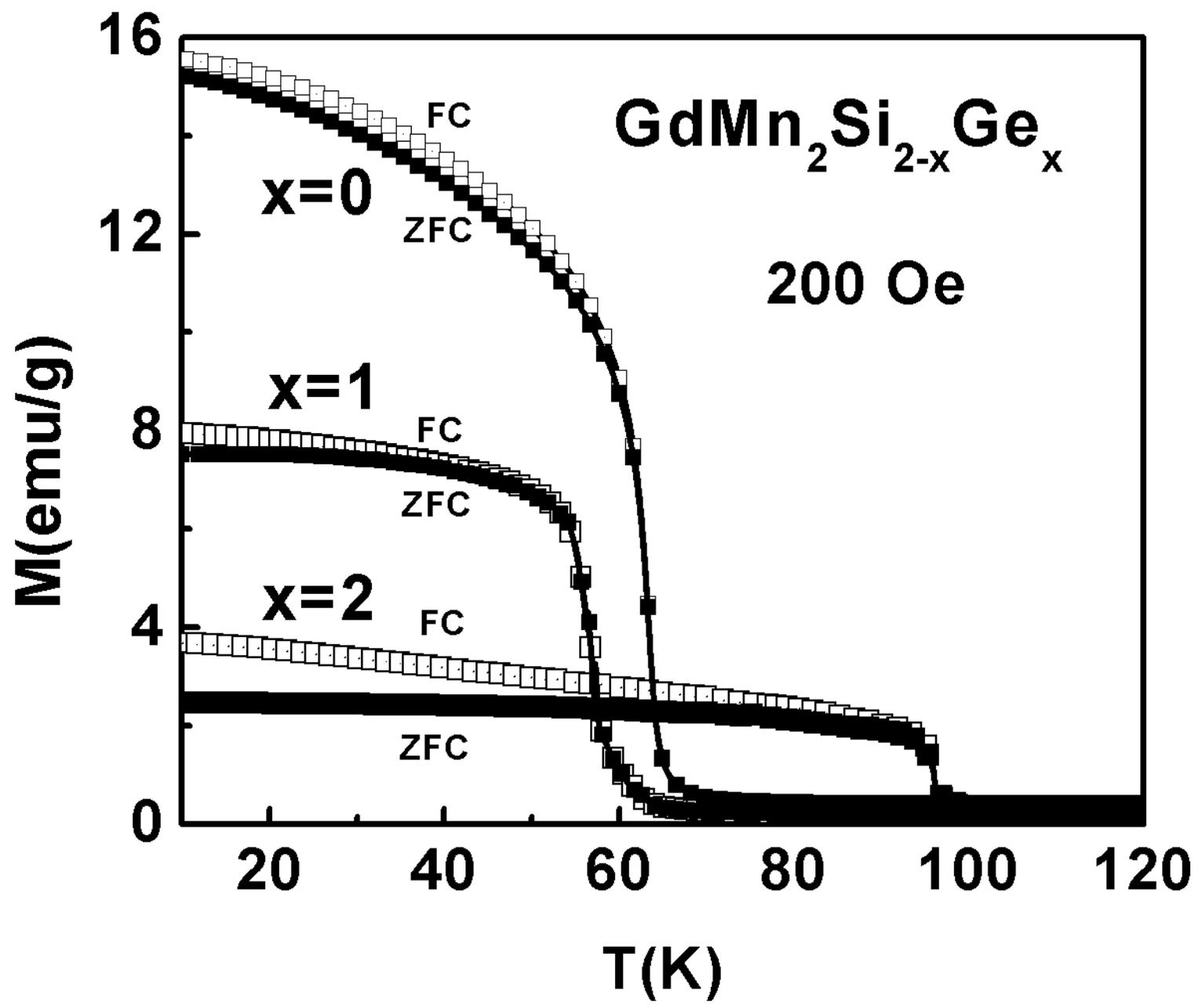

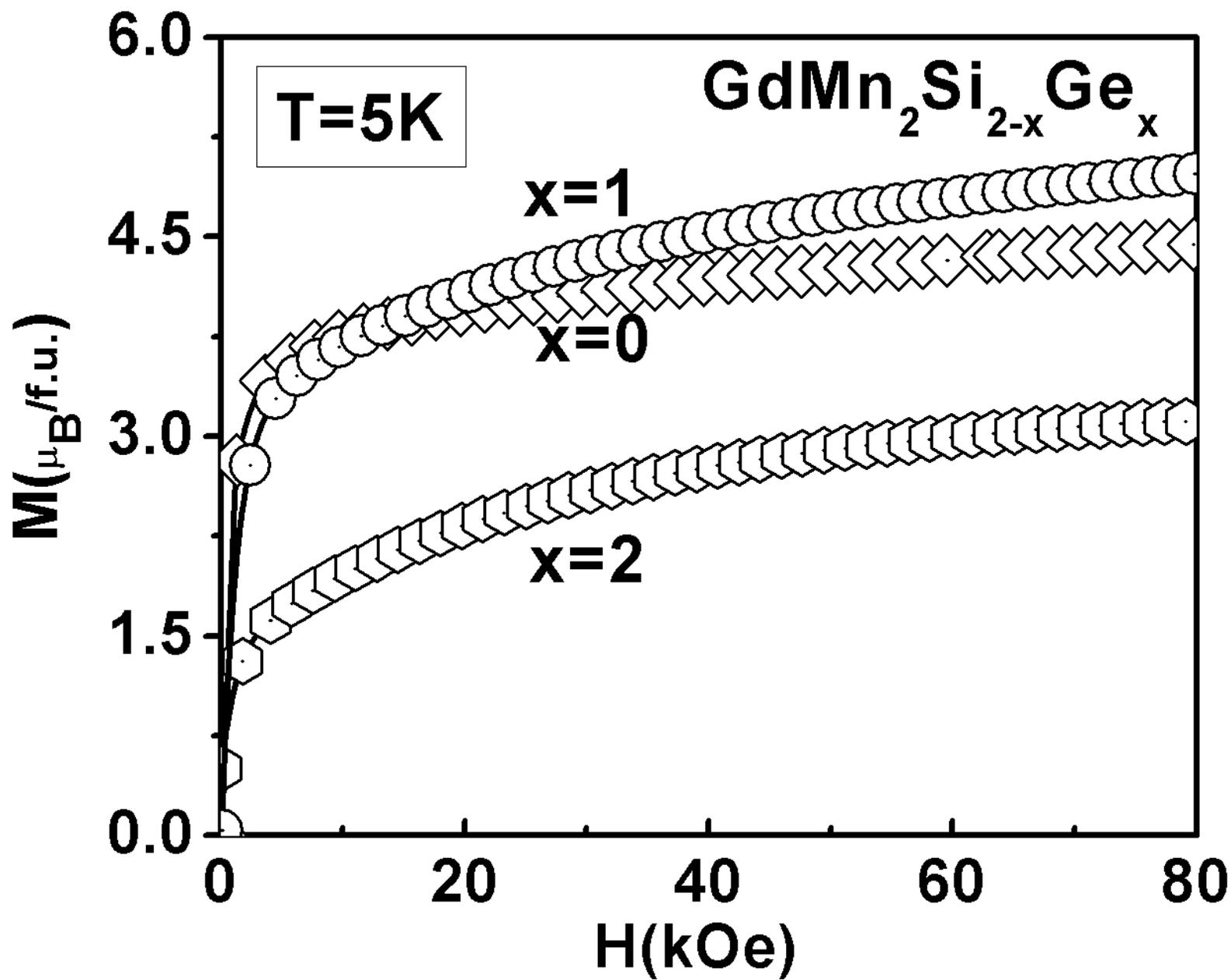

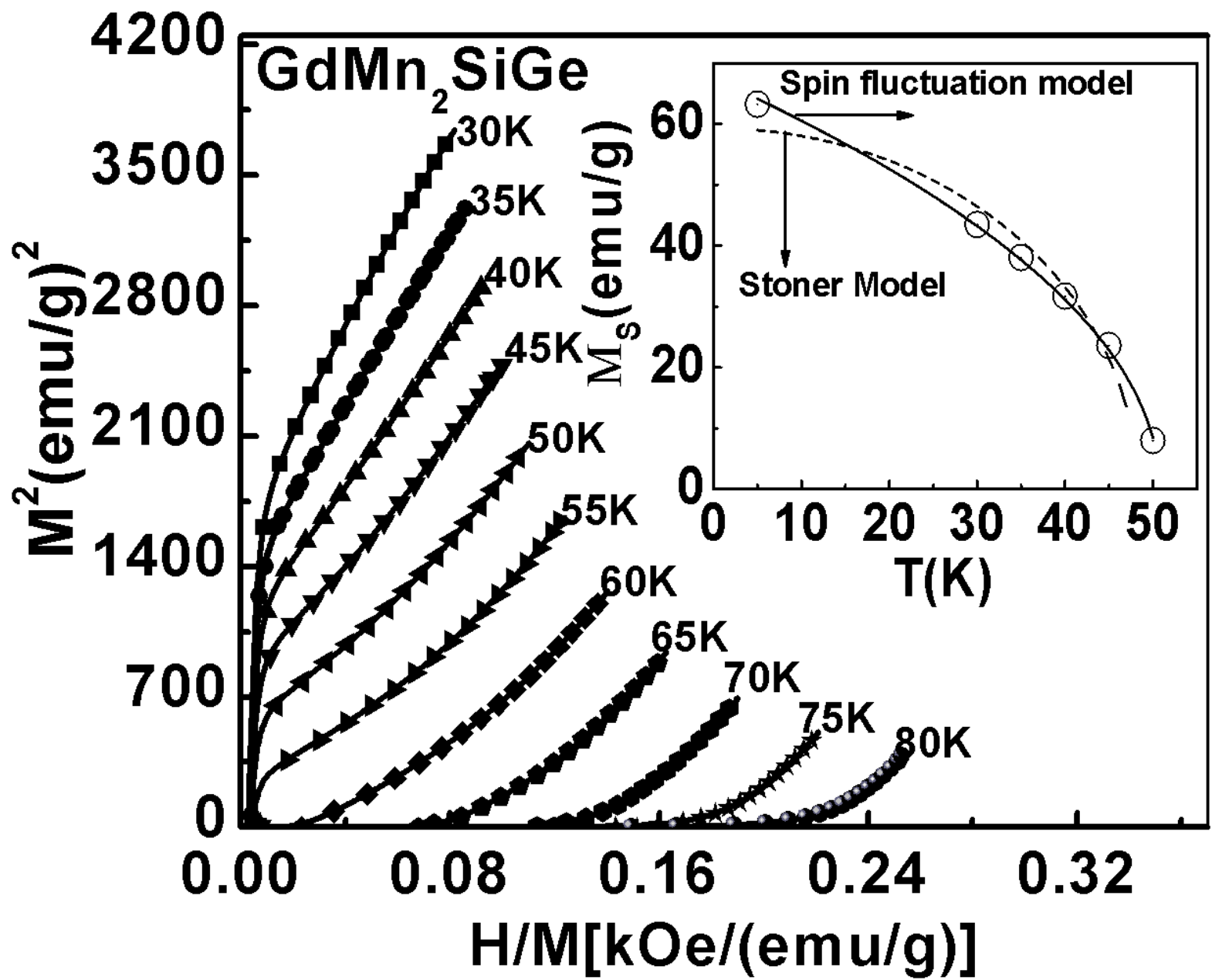

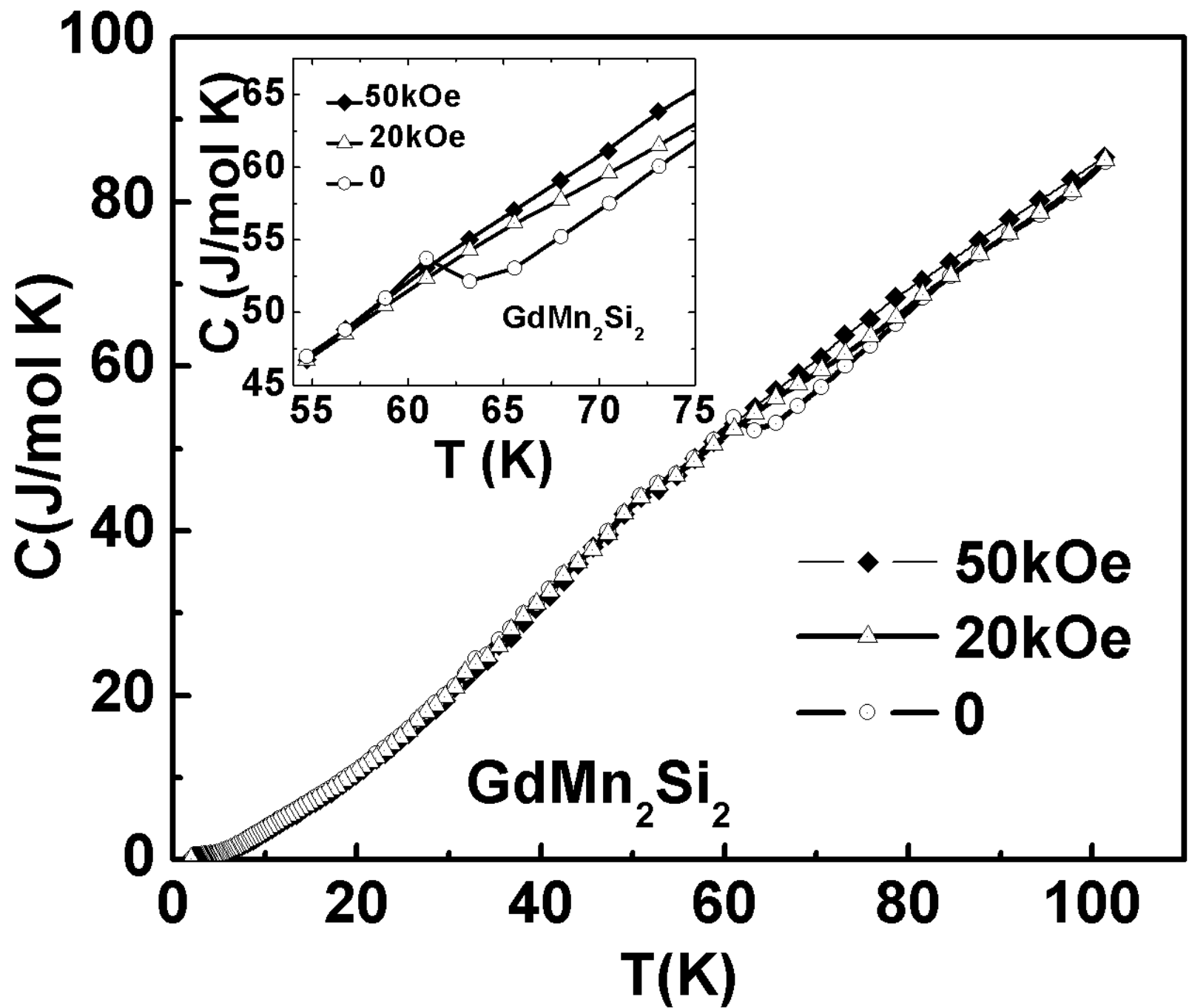

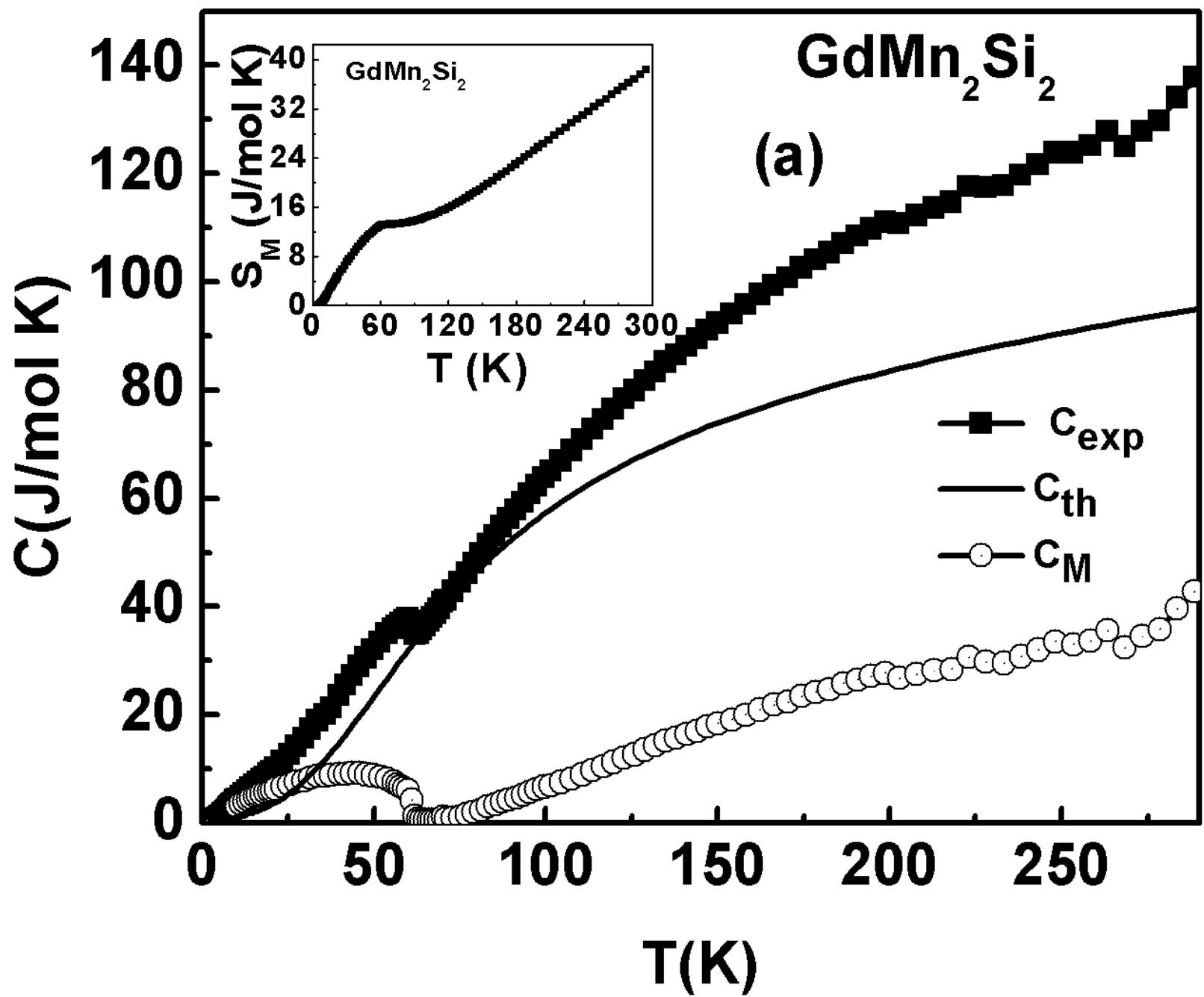

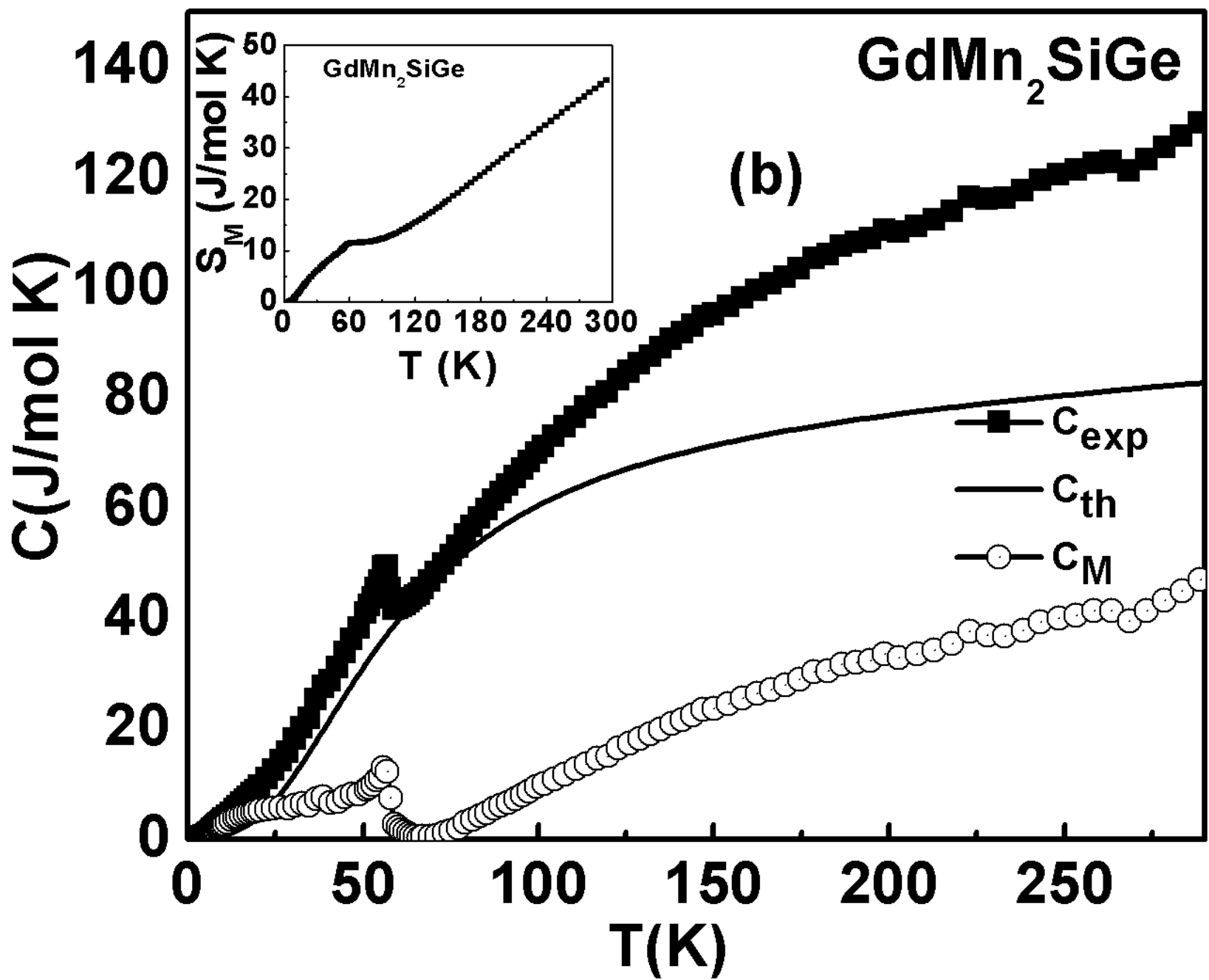

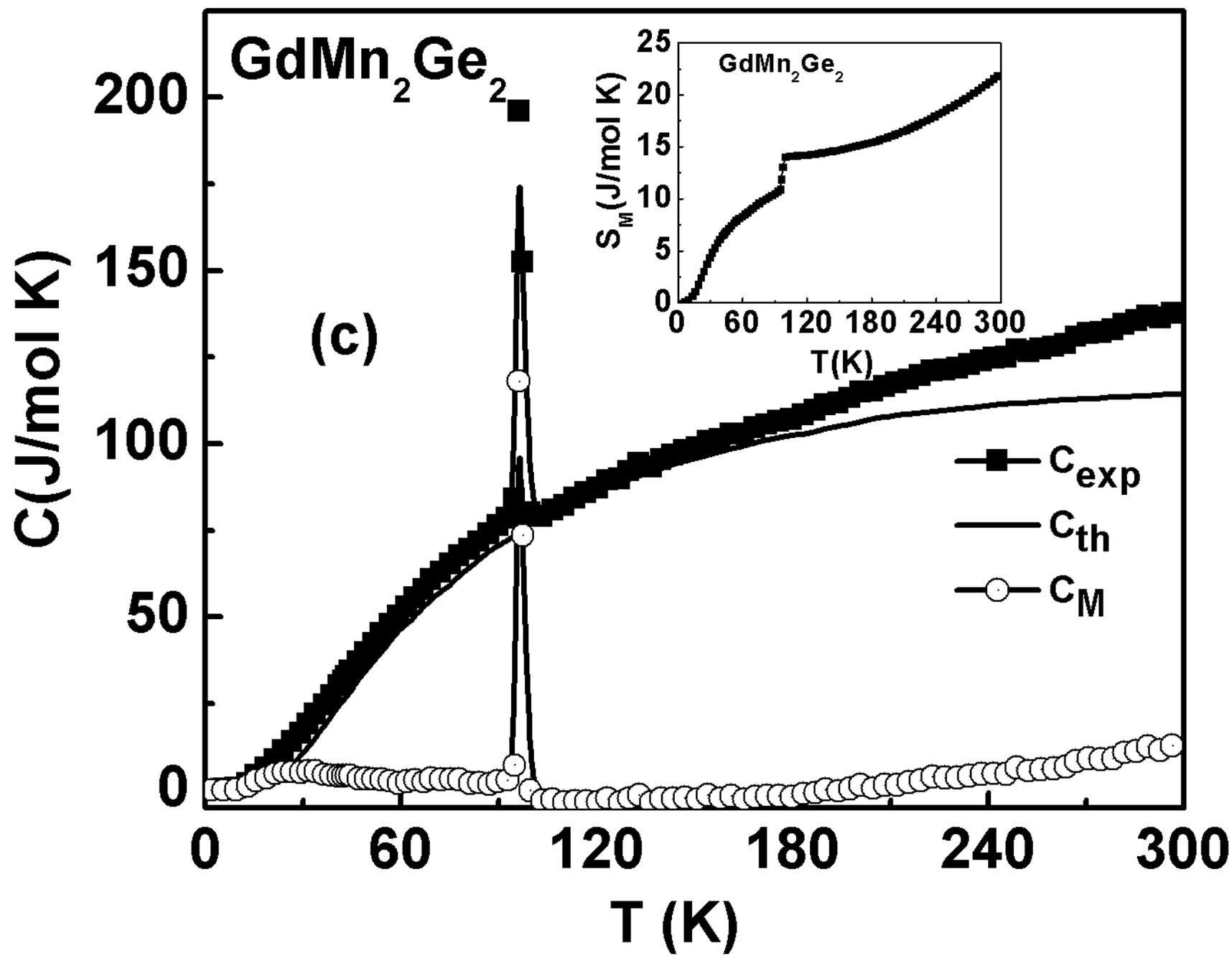

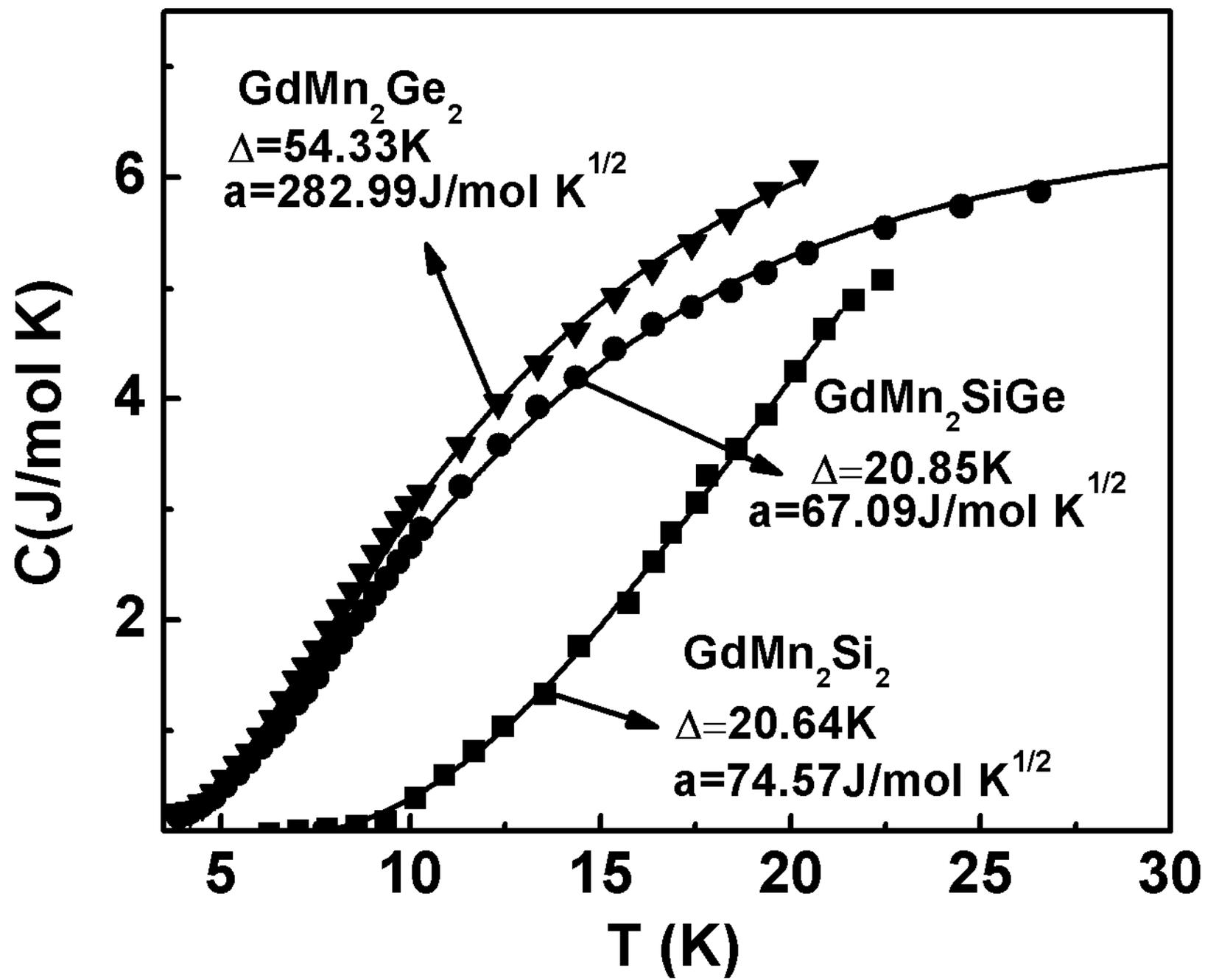

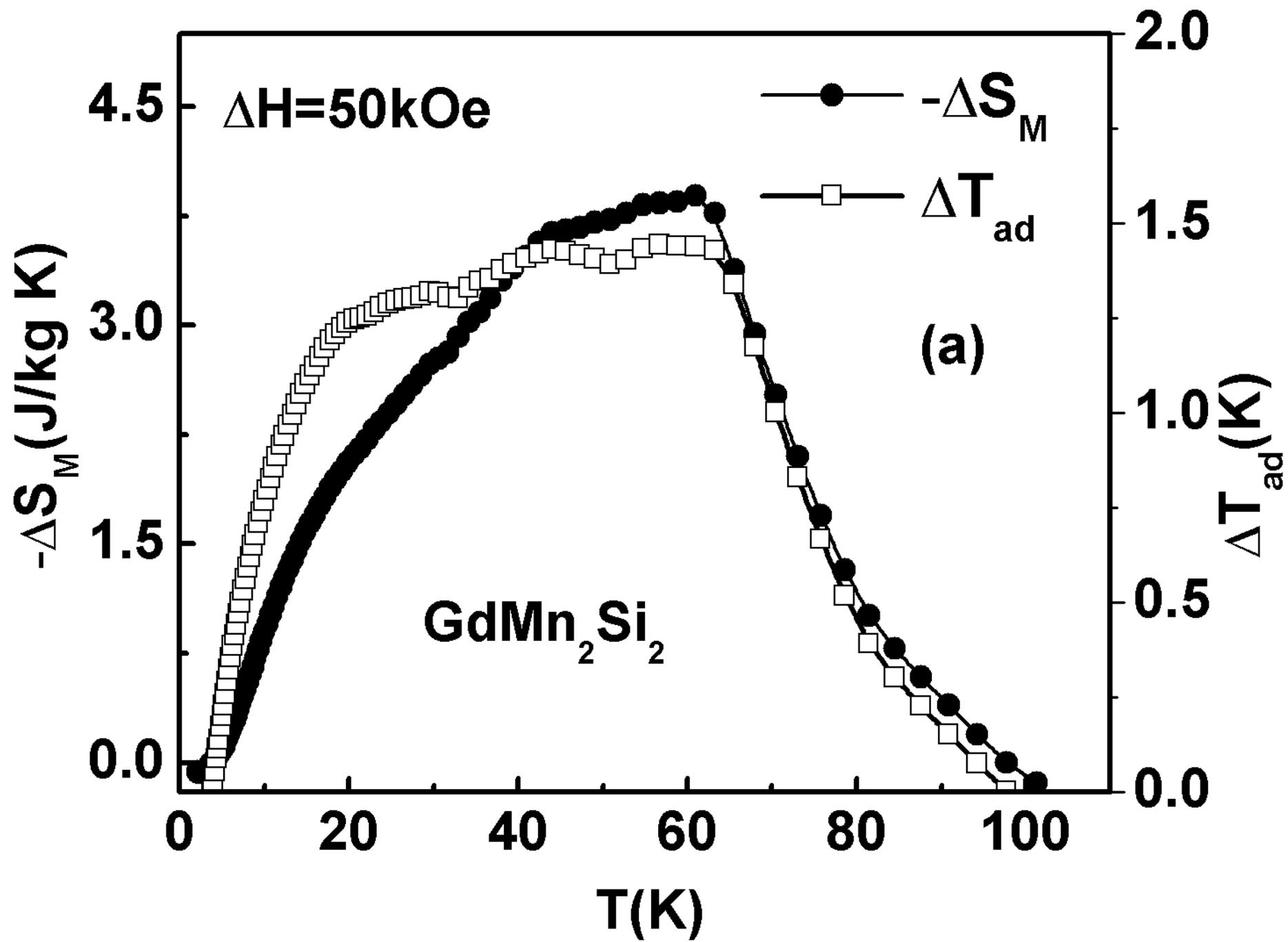

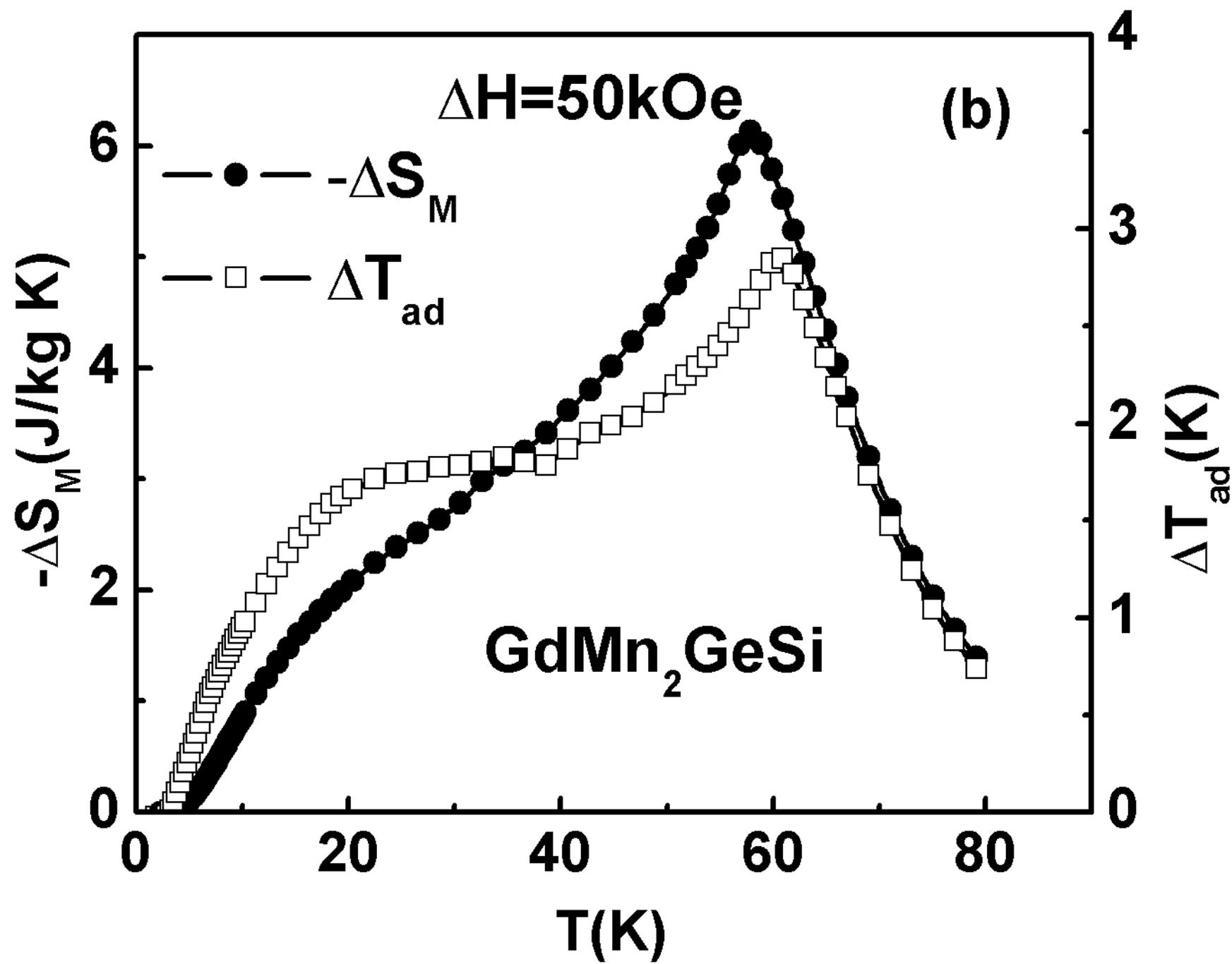

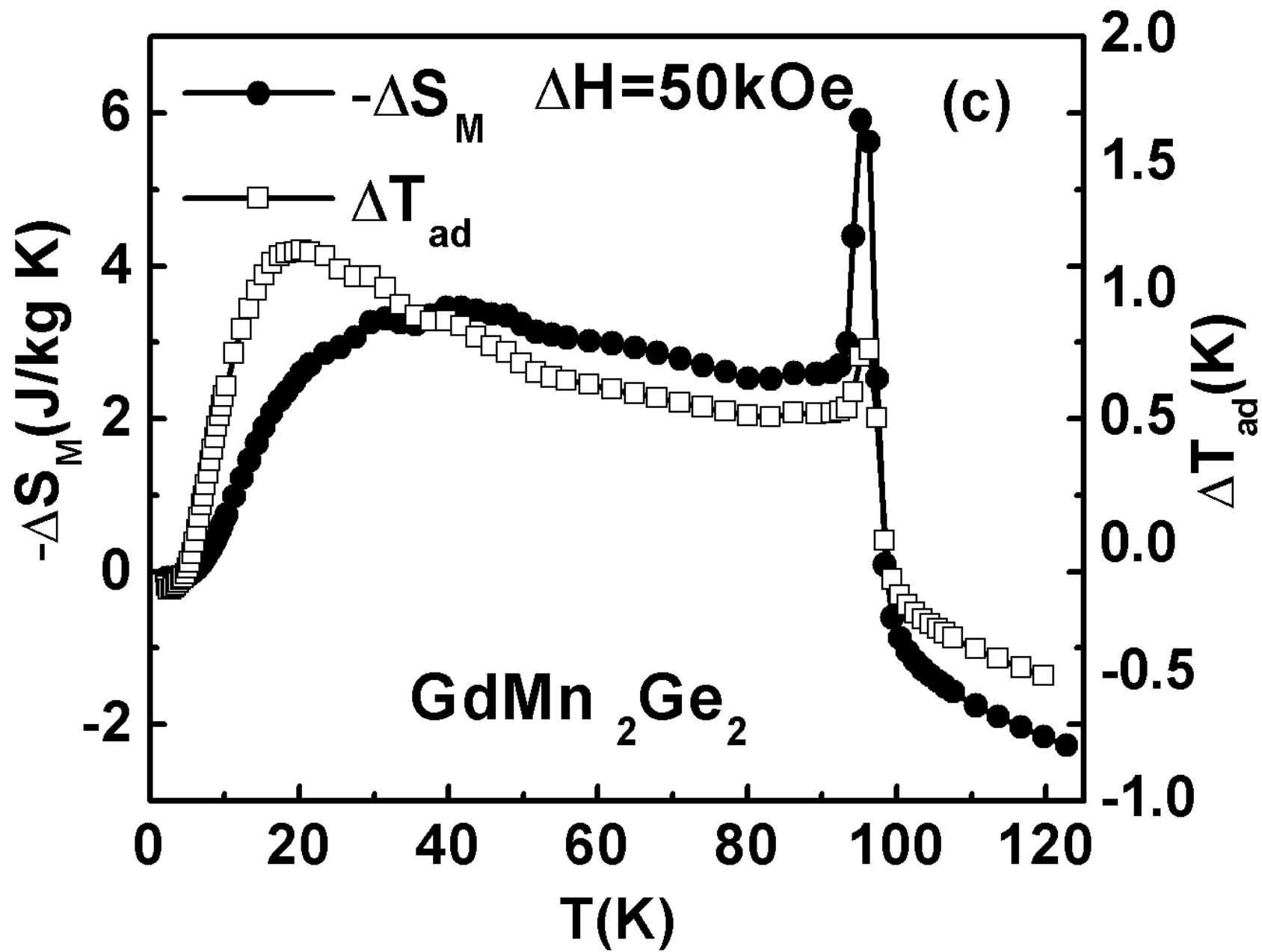

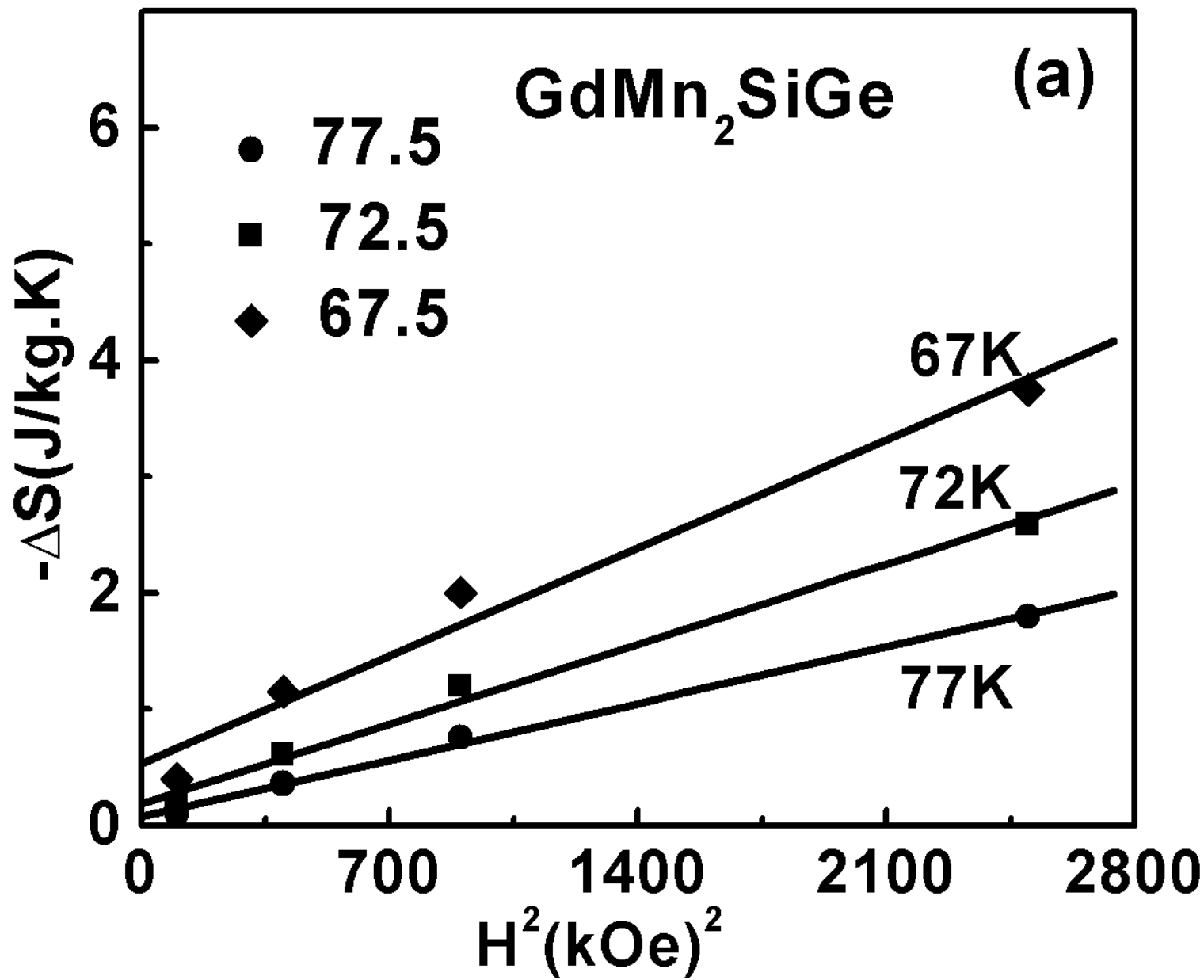

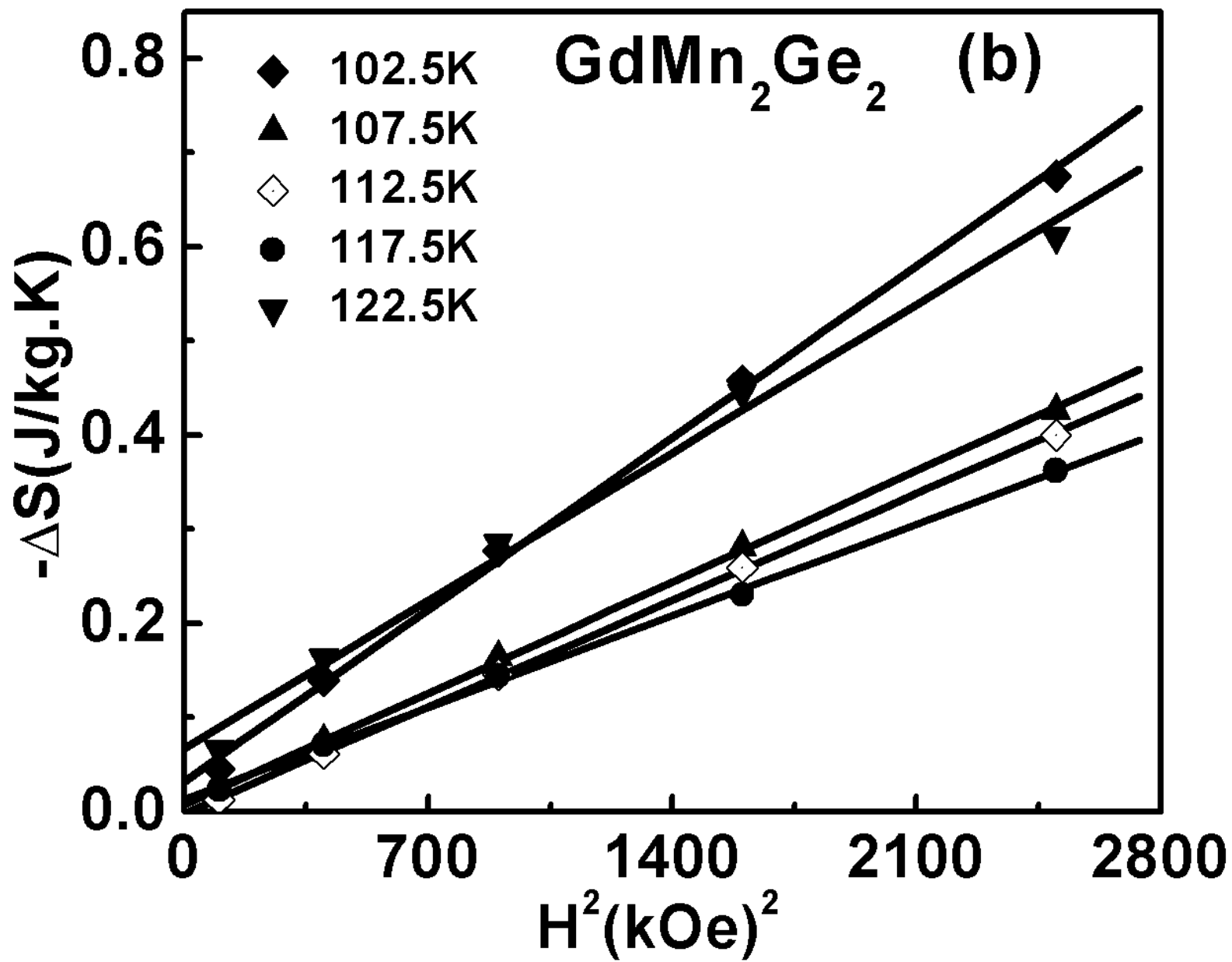

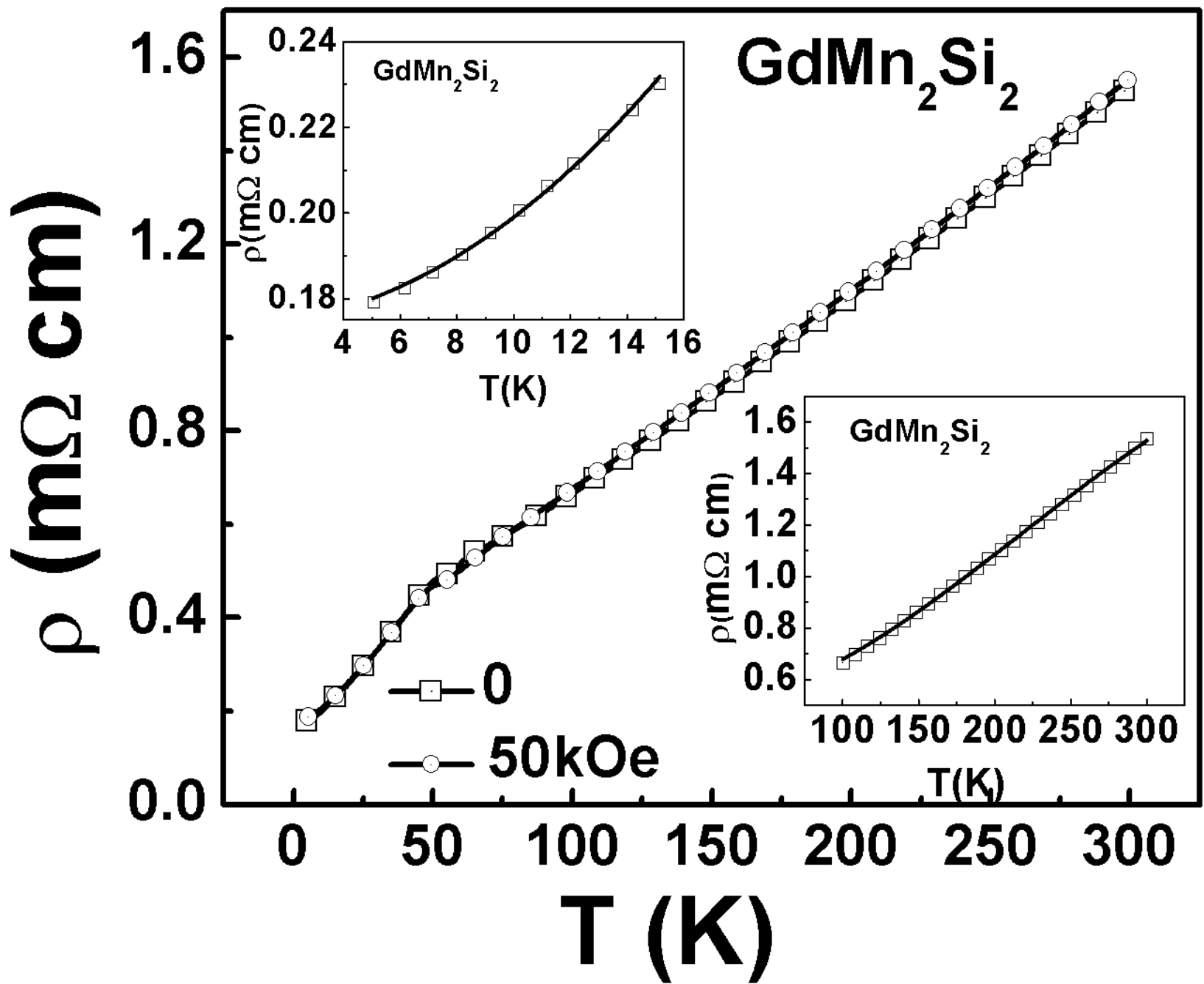

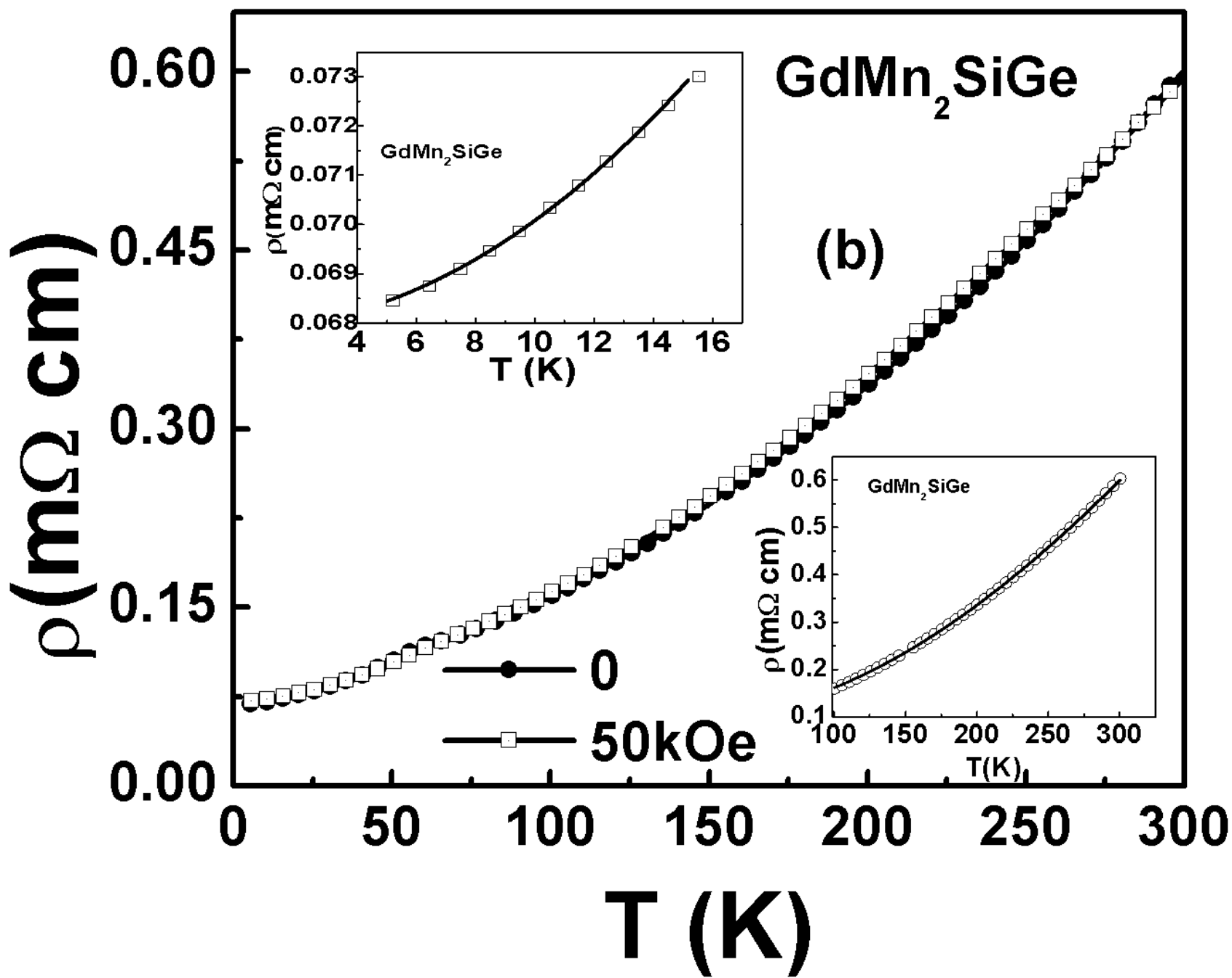

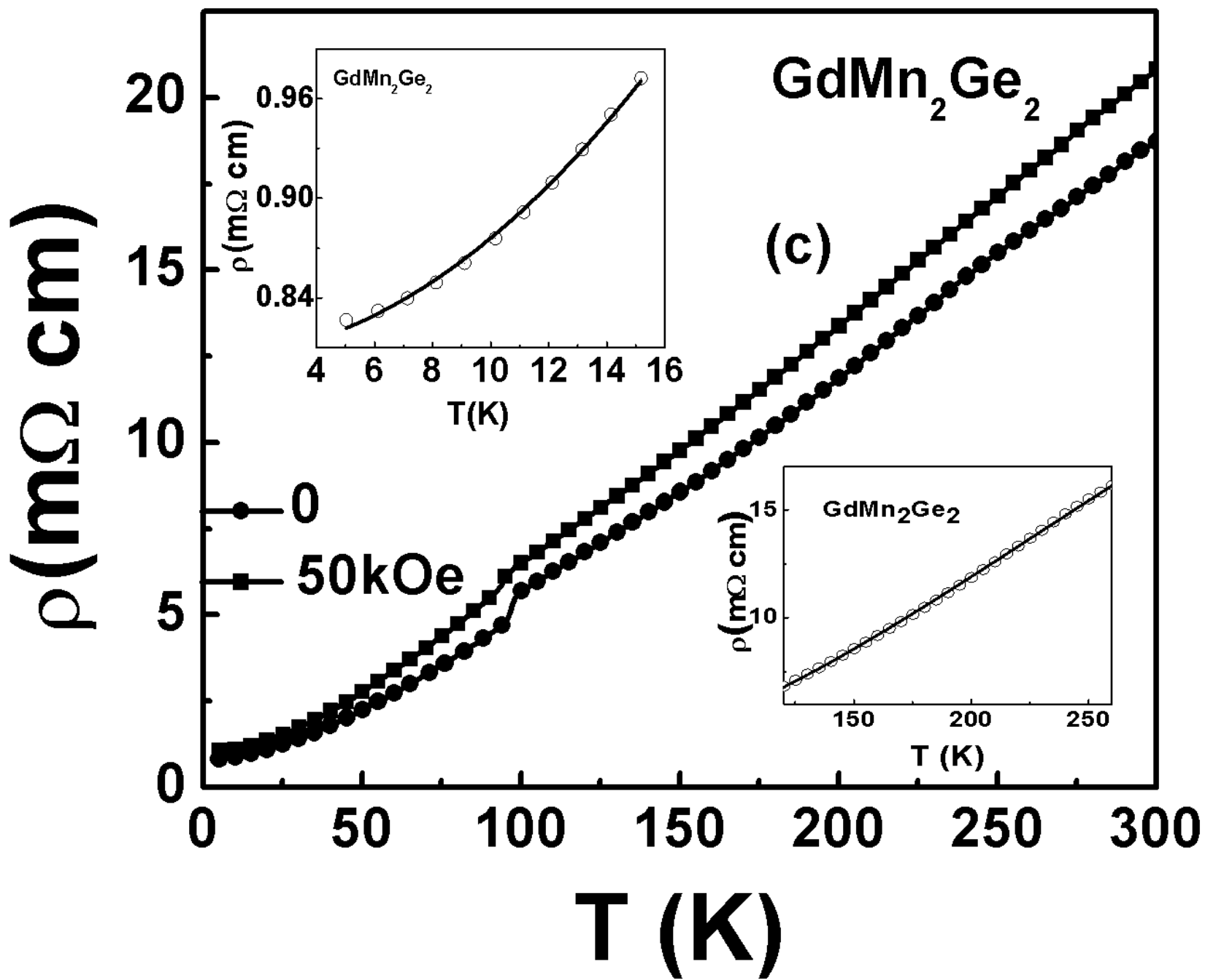

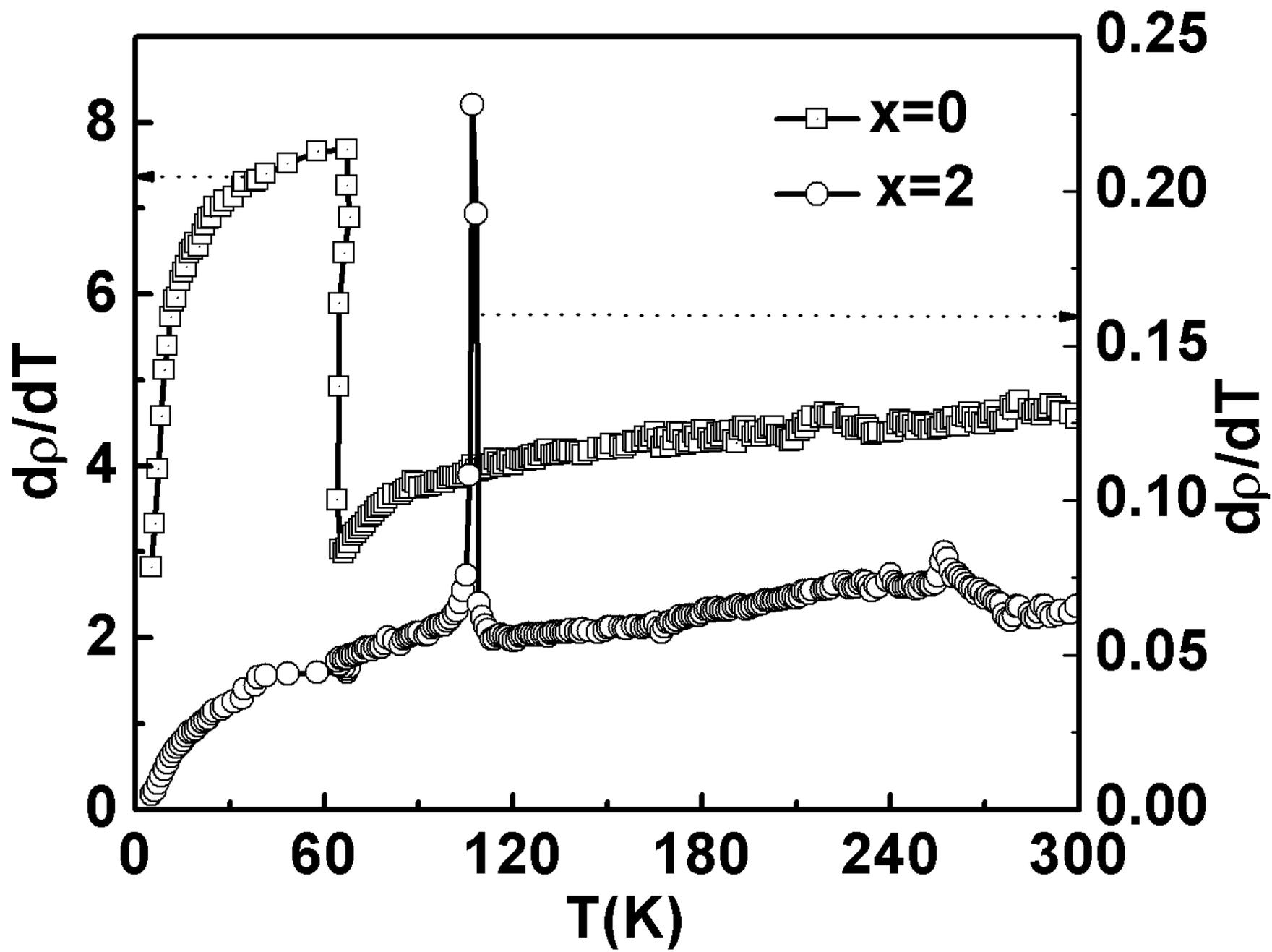

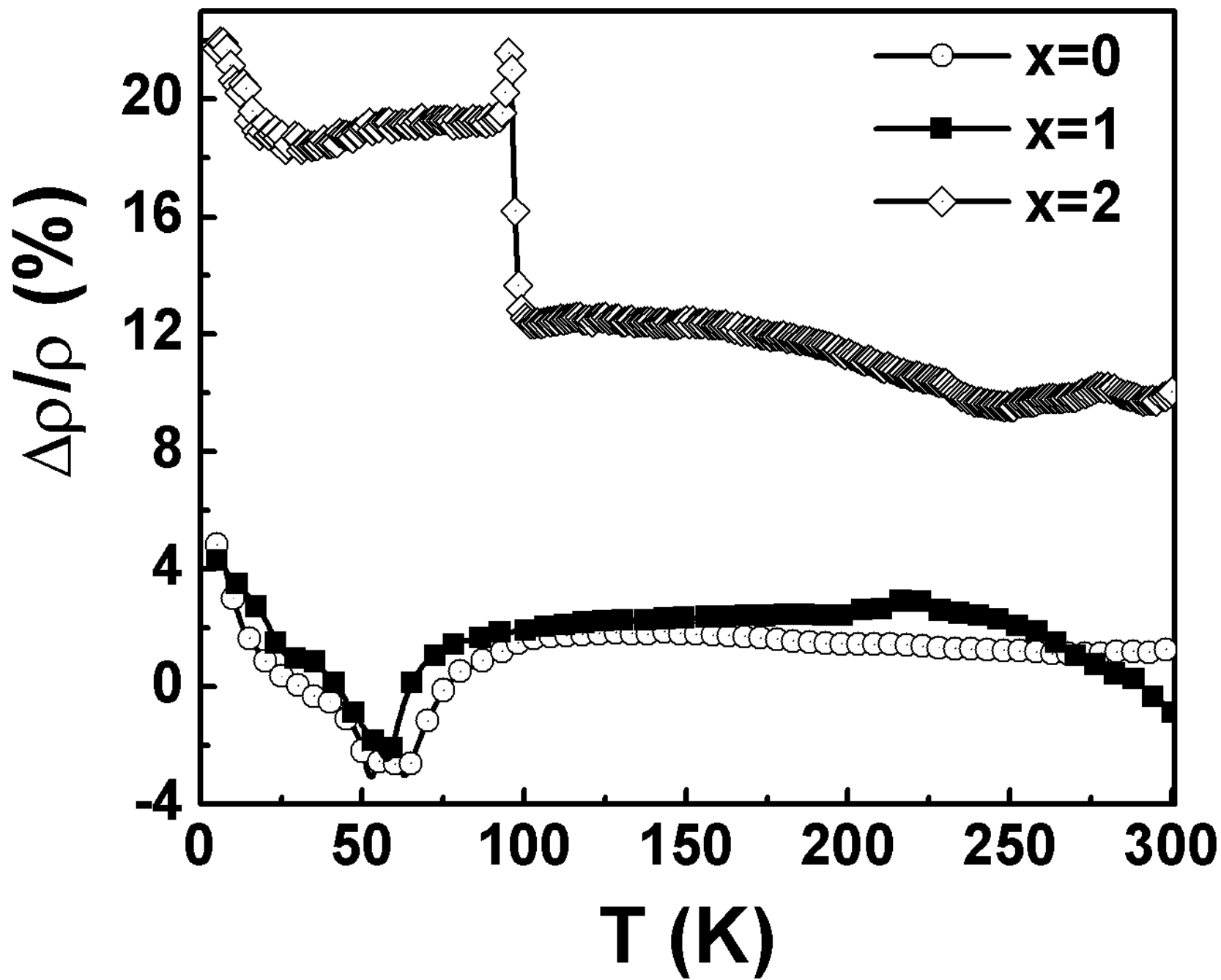